\documentclass[prb,aps,twocolumn]{revtex4-2}%

\usepackage{amsmath}
\usepackage[utf8]{inputenc}
\usepackage{csquotes}
\usepackage{graphicx} %
\usepackage{amssymb}
\usepackage{verbatim}
\usepackage[colorlinks=true, pdfstartview=FitV, linkcolor=blue, citecolor=blue, urlcolor=blue]{hyperref} %
\usepackage{pifont}
\usepackage{rotating}

\def\C        {{$^{13}$C \/}}

\def\NN       {{$^{15}$N \/}}

\def\eg       {{\it e.g.}}

\newcommand{\mr}[1]{\mathrm{#1}}
\newcommand{\unit}[1]{\,\mathrm{#1}}

\newcommand{\us}{\,\mu{\rm s}}

\newcommand{\yn}{\gamma_\mr{n}}

\newcommand{\ket}[1]{\ensuremath{\left|#1\right\rangle}}

\newcommand{\apar}{a_{||}}
\newcommand{\Dapar}{\Delta a_{||}}
\newcommand{\aperp}{a_{\perp}}

\newcommand{\wD}{\omega_\mr{D}}
\newcommand{\wo}{\omega_0}

\begin{document}

\title{Multidimensional spectroscopy of nuclear spin clusters in diamond}

\author{Konstantin~Herb$^{1}$, Takuya~F.~Segawa$^{1,2}$, Laura~A.~V\"olker$^{1}$, John~M.~Abendroth$^{1}$, Erika~Janitz$^{1,3}$, Tianqi~Zhu$^{1}$, and Christian~L.~Degen$^{1,4}$}
\email{degenc@ethz.ch}
\affiliation{$^1$Department of Physics, ETH Zurich, Otto Stern Weg 1, 8093 Zurich, Switzerland.}
\affiliation{$^2$Department of Chemistry and Applied Biosciences, ETH Zurich, Vladimir Prelog Weg 1-5/10, 8093 Zurich, Switzerland.}
\affiliation{$^3$Department of Electrical and Software Engineering, University of Calgary, Calgary Alberta T2N 1N4, Canada}
\affiliation{$^4$Quantum Center, ETH Zurich, 8093 Zurich, Switzerland.}

\begin{abstract}
\noindent
Optically active spin defects in solids offer promising platforms to investigate nuclear spin clusters with high sensitivity and atomic-site resolution.  To leverage near-surface defects for molecular structure analysis in chemical and biological contexts using nuclear magnetic resonance (NMR), further advances in spectroscopic characterization of nuclear environments are essential.
Here, we report Fourier spectroscopy techniques to improve localization and mapping of the testbed $^{13}\mathrm{C}$ nuclear spin environment of individual, shallow nitrogen-vacancy centers at room temperature.
We use multidimensional spectroscopy, well-known from classical NMR, in combination with weak measurements of single-nuclear-spin precession.
We demonstrate two examples of multidimensional NMR:  (i) improved nuclear spin localization by separate encoding of the two hyperfine components along spectral dimensions and (ii) spectral editing of nuclear-spin pairs, including measurement of internuclear coupling constants.
Our work adds important tools for the spectroscopic analysis of molecular structures by single-spin probes.
\end{abstract}

\maketitle

Electronic spin defects in wide-bandgap dielectrics, including diamond, silicon carbide, or oxide crystals, provide exceptionally long spin coherence times, even at room temperature~\cite{atature18}.  Combined with efficient optical spin preparation and readout, such defect spins can be used to analyze local nuclear spin environments~\cite{abobeih19} and conduct NMR at the single-spin level~\cite{jelezko04nuclear,childress06,cujia19}. Particularly, nitrogen-vacancy (NV) defect spins in diamond have been used to detect and control individual spins in clusters comprising multiple nuclei, and offer potential routes to resolve the structure of proximal molecules with atomic resolution and elemental specificity~\cite{degen08apl,ajoy15,kost15,perunicic16}.

The naturally occurring \C spin bath in diamond offers an ideal testbed for developing single-spin imaging techniques. Early work employed dynamical decoupling (DD) spectroscopy to detect \C resonances corresponding to individual nuclei~\cite{zhao12,taminiau12,kolkowitz12,bradley19}.  Spectral assignment can be used to determine spatial positions or identify spin pairs \cite{kalb16,bartling22}.  However, the spectral resolution of DD is limited by NV coherence time $T_2$, and restricted to a few kHz unless under cryogenic conditions~\cite{abobeih18}.  More recently, alternative methods such as correlation spectroscopy~\cite{laraoui13} and weak-measurement schemes~\cite{pfender19,cujia19} have improved spectral resolution to the Hz range at room temperature.

These advances have enabled atomic-scale localization of \C spins in the diamond lattice~\cite{zhao11, zhao12, sasaki18, zopes18ncomm, zopes18prl}.  To date, clusters of roughly 20 nuclei at room temperature~\cite{cujia22} and 50 nuclei at cryogenic temperatures~\cite{stolpe23} have been resolved.  Despite this progress, mapping large nuclear spin networks remains challenging due to spectral overlap, positional ambiguities and internuclear interactions~\cite{cujia22}.  Multidimensional techniques have proven to reduce such ambiguities at cryogenic temperatures~\cite{abobeih19,stolpe23}, however, translating these approaches to ambient conditions is not straightforward because of reduced spin lifetimes.

In this work, we combine multidimensional Fourier spectroscopy with weak measurements of nuclear spin precession to establish a general route to improving the accuracy of spin localization.  First, we present correlative mapping of the transverse and parallel hyperfine coupling parameters~\cite{boss16} to triangulate global nuclear spin positions. Second, we edit nuclear spin-spin couplings using a variant of $J$-spectroscopy~\cite{aue76} to identify spin-pair configurations.  In both cases, 2D spectra remove overlap present in 1D spectra and enable improved assignment of single nuclear peaks.
We demonstrate our schemes on the testbed \C environment of shallow NVs at room temperature, which is compatible with future experiments on surface-anchored molecules.
Further work on stabilizing NV centers near the diamond surface~\cite{sangtawesin19,janitz22} and developing reliable chemical functionalization of the surface~\cite{liu22,xie22,abendroth22} will be needed to realize this exciting potential.

\begin{figure*}
	\centering
	\includegraphics[width=\textwidth]{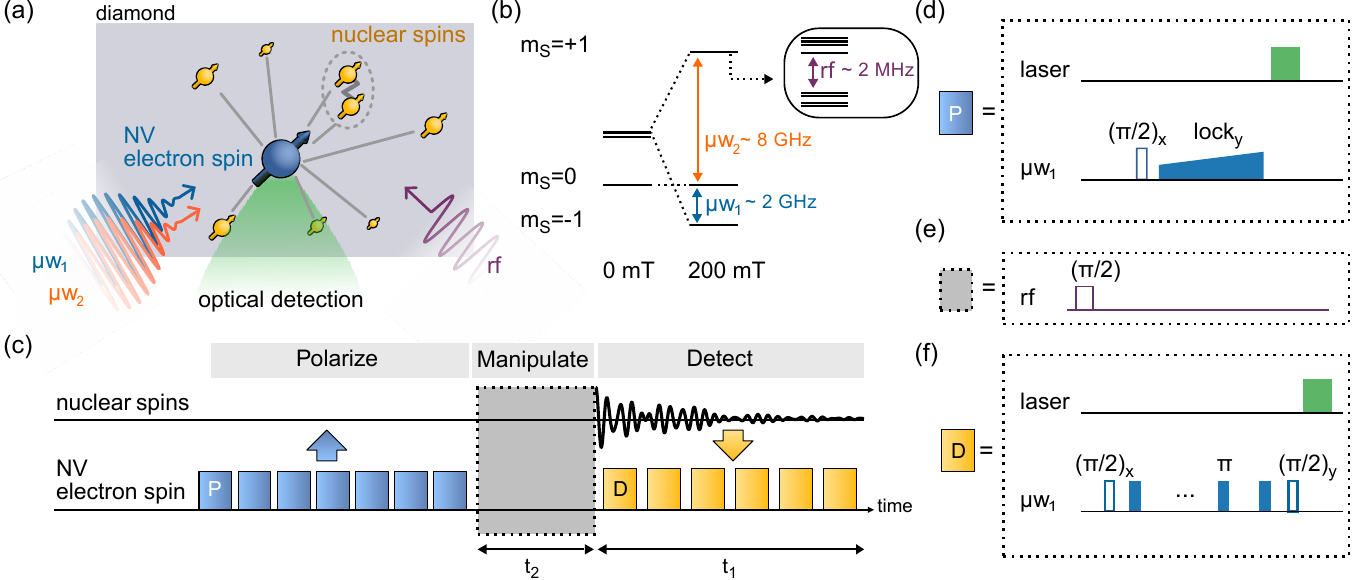}
	\caption{
  (a) Experiment schematic. An NV electron spin (blue) interacts with proximal \C nuclear spins (yellow) in the diamond (gray) \textit{via} the dipolar hyperfine interaction.  Optical preparation and readout of the NV spin state is achieved using laser pulses and photoluminescence intensity measurements.  Arrows indicate microwave ($\mu$w; blue, orange) and radio-frequency (rf; purple) tones for electron and nuclear spin manipulation, respectively.
  (b) Energy-level diagram for the NV ground-state manifold, including the bath of \C nuclei (inset).  The \NN\ spin is neglected for clarity.
  (c) Canonical sequence for 1D NMR spectroscopy using the weak-measurement detection scheme.
  (d-f) Detail of the ``polarize'', ``manipulate'' and ``detect'' blocks. 
  In (d-f), green denotes laser pulses, blue denotes microwave pulses ($\mu$w$_1$), and purple denotes the rf pulse. Horizontal axis is time.
  }
	\label{fig1}
\end{figure*}

Our experimental scheme is shown in Fig.~\ref{fig1}a.  A central electronic NV spin is surrounded by a dilute bath ($\sim 1\%$ natural abundance) of \C nuclear spins.  NVs are created by ion implantation in a single-crystalline diamond chip. Subsequent annealing results in shallow ($<10\unit{nm}$-deep \cite{abendroth22}) NVs.  Following defect creation, nanopillar waveguides are fabricated to enhance optical collection efficiency~\cite{zhu23}.  Off-resonant (532~nm) laser pulses ($\sim2.5\us$ duration) are used to optically polarize and read out the spin state.  A two-channel microwave generator connected to a broadband co-planar waveguide antenna serves to drive the two allowed electronic spin transitions at frequencies $\mu$w$_1$ and $\mu$w$_2$ (Fig.~\ref{fig1}(b)).
Typical $\pi$-pulse durations are $20\unit{ns}$ ($80\unit{ns}$) for $\mu$w$_1$ ($\mu$w$_2$), respectively .  Radio-frequency (rf) pulses for \C manipulation are applied \textit{via} a separate broadband microcoil antenna~\cite{herb20}, with a typical $\pi$-pulse duration of $8\unit{\us}$.  Further details are given in the SM~\cite{supplemental}.

Fig.~\ref{fig1}(c-f) revisit the protocol for 1D weak-measurement spectroscopy~\cite{cujia19,cujia22}.  The basic 1D protocol consists of three steps:
First, the NV spin is polarized by a laser pump pulse followed by transfer of the NV polarization to the nuclei using the nuclear spin orientation \textit{via} electron spin locking (NOVEL) protocol~\cite{henstra88} (Fig.~\ref{fig1}(d)).
Since one transfer cycle can polarize a single \C at most, this cycle is repeated $\sim 10^3$ times to polarize a nm-scale volume of \C around the NV~\cite{cujia22}.  Second, a single rf $\pi/2$ pulse triggers nuclear precession (Fig.~\ref{fig1}e).  Third, precessing nuclear magnetization is detected \textit{via} a series of weak measurements (Fig.~\ref{fig1}f) \cite{colangelo17,pfender19,cujia19}. The overall sequence is averaged over $\sim 10^6$ cycles to obtain sufficient signal-to-noise ratios. Resulting time-domain signals reflect the free-induction decay signal of nuclear precession~\cite{cujia19}.  Fourier transformation with respect to the free precession time $t_1$ then yields the 1D NMR spectrum of the NV's \C environment~\cite{cujia22}.

\begin{figure*}
	\centering
	\includegraphics[width=1.00\textwidth]{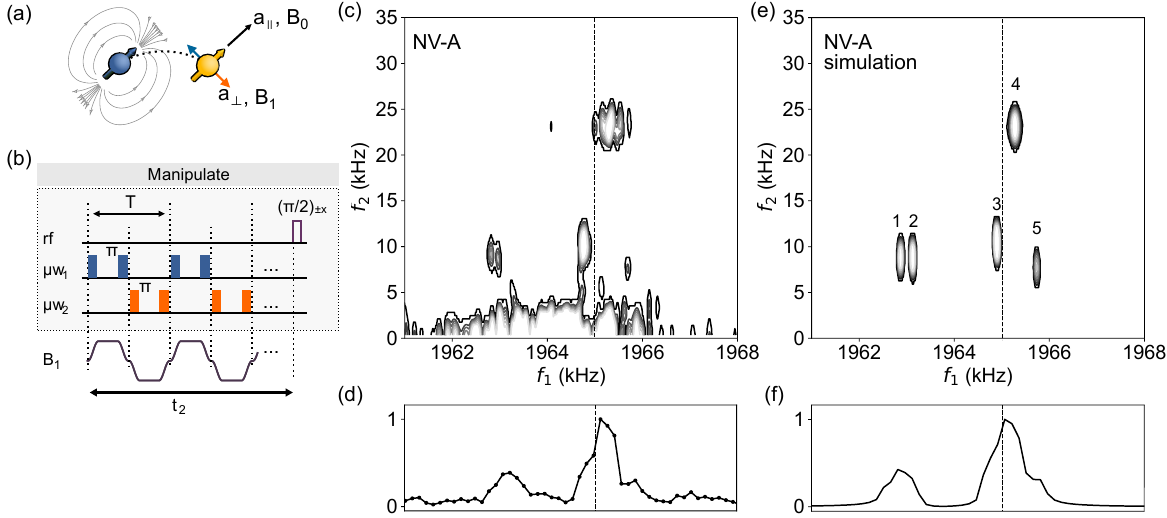}
	\caption{Two-dimensional hyperfine spectroscopy.
	(a) Periodic inversion of the NV spin leads to modulation of the hyperfine interaction. Blue (orange) shows the direction of the transverse hyperfine coupling parameter $\aperp$ when the NV spin is in the $\ket{-1}$ ($\ket{+1}$) state.
	(b) Corresponding pulse sequence. Periodic inversion of the NV spin creates an effective hyperfine magnetic field of amplitude $B_1$ experienced by \C spins. The modulation period $T=2\pi/\wo$ corresponds to one nuclear Larmor period, where $\wo = \yn B_0$.
	(c) 2D Fourier power spectrum with respect to $t_1$ and $t_2$.  $f_1$ and $f_2$ are the corresponding frequency axes. The horizontal axis $f_1$ encodes the parallel hyperfine component $\apar$, while the vertical axis $f_2$ encodes the transverse hyperfine component $\aperp$ (Eq.~\ref{eq:freq_convert}).  The ``bare'' Larmor frequency $\wo/2\pi$ is indicated by a dashed vertical line.  The time increment and number of points are $11.36\unit{\us}$ and $1925$ for the $t_1$ dimension, and $11.2\unit{\us}$ ($=22T$) and $126$ for the $t_2$ dimension, respectively.
	(d) 1D projection of the 2D Fourier spectrum.
	(e,f) Numerical simulation of the five main peaks using the density matrix method.
  }
	\label{fig2}
\end{figure*}

1D spectroscopy of the \C nuclear spin bath has been used to determine spatial locations of nuclei surrounding NVs~\cite{zhao11,taminiau12,zopes18ncomm,cujia22}. Spatial mapping is achieved by analyzing the parallel and transverse components of the hyperfine interaction tensor, denoted by $\apar$ and $\aperp$, which uniquely define the radial distance $r$ and polar angle $\theta$ between the two spins (up to an inversion at the origin)~\cite{boss16}.  Analysis of the nuclear precession phase permits extracting the azimuth $\phi$~\cite{sasaki18,zopes18prl}.  Together, the two measurements yield the full 3D location vector ($r$,$\theta$,$\phi$) of a nuclear spin.  While this approach has been successfully extended to several nuclear spins, mapping of larger clusters remains challenging because of spectral overlap and the presence of internuclear couplings. Further, unequal polarization of nuclear spins and varying local relaxation environments contribute to localization errors~\cite{cujia22}.  Therefore, 1D methods become imprecise for larger spin clusters ($n\gtrsim 10$), necessitating more advanced methods for constraining positions.

In the following, we show that multidimensional spectroscopy can address some ambiguities and limitations associated with 1D methods.  While multidimensional approaches have been considered in the context of NVs~\cite{boss16,abobeih19}, we take inspiration from 2D~NMR~\cite{aue76} and show that weak-measurement spectroscopy can be extended to multiple spectral dimensions by introducing an indirect evolution time $t_2$ during the ``manipulation'' block (Fig.~\ref{fig1}(c)). Using this approach, we demonstrate correlative measurement of hyperfine coupling constants and editing of dipolar-coupled spin pairs.

Fig.~\ref{fig2} presents a first example of multidimensional spectroscopy, where we correlate the two coupling parameters ($\apar$,$\aperp$) of the hyperfine tensor.  The goal of this sequence is to improve the accuracy of the $\aperp$ measurement and resolve overlapping peaks with similar $\apar$ in 1D spectra.  Specifically, we replace the $\pi/2$ rf pulse of the ``manipulation'' period (Fig.~\ref{fig1}(e)) with a periodic inversion of the NV spin commensurate with the \C Larmor precession (Fig.~\ref{fig2}(b)).  Periodic nuclear inversion results in modulation of the hyperfine coupling, which in turn causes Rabi driving of the nuclear spins.  The amplitude of the driving field is given by 
\begin{align}
B_1 = \frac{4\aperp}{\pi\yn} \left(1 - \left({\pi^2-4}\right) \frac{t_\pi^2}{T^2} \right)\text{,}
\end{align}
where $t_\pi$ is the duration of the electronic $\pi$ pulse, $T$ is the modulation period, and $\yn = 2\pi\times 10.71\unit{MHz/T}$ is the \C gyromagnetic ratio.
In contrast to previous Rabi driving experiments on single nuclei~\cite{taminiau14,boss16}, inversion of both electronic spin transitions is essential to drive nuclear ensembles.  Because each nuclear spin experiences different static detunings by the $\apar$ term, inverting only one NV transition leads to large dispersion of Rabi frequencies due to off-resonant driving. Note that inverting both NV transitions doubles the $B_1$ amplitude (Fig.~\ref{fig2}(b)).

To obtain 2D spectra, we acquire free-induction decays for a series of $t_2$ increments, where $t_2$ is the duration of the periodic inversion block.  Applying 2D Fourier transformation with respect to $t_1$ and $t_2$ generates the desired 2D NMR spectrum.

\begin{figure*}
	\centering
	\includegraphics[width=\textwidth]{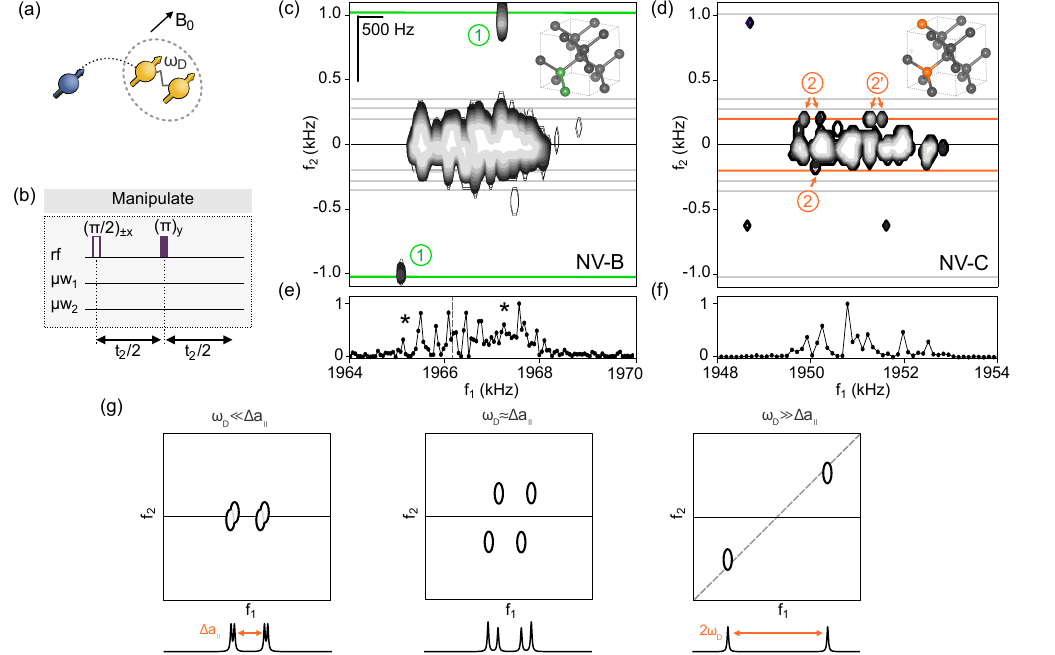}
	\caption{Two-dimensional $J$-spectroscopy of nuclear spin pairs. 
	(a) Illustration of an NV spin coupled to a nuclear spin pair.
	(b) Pulse sequence consisting of a nuclear spin echo formed by two rf $\pi/2$ and $\pi$ pulses, respectively.  The single echo could be extended to multiple echoes using a CPMG sequence. Indirect evolution time $t_2$ is incremented in steps of $300\unit{\us}$ (NV-B) and $388\unit{\us}$ (NV-C), respectively.  No microwave pulses are applied.
	(c,d) 2D Fourier spectra for NV-B and NV-C. The horizontal axis $f_1$ encodes the full free precession Hamiltonian. The vertical axis $f_2$ only encodes nuclear spin-spin coupling $\wD$ (Eq.~\ref{eq:wD}).  Horizontal lines indicate frequencies of expected spin-pair configurations.  Atomic arrangements for identified pairs (1), (2) and (2') are shown as insets.
	(e,f) 1D projections of the respective 2D spectra.  In (e), satellite peaks are marked by asterisks.
	(g) Expected peak patterns for a spin pair with weak, intermediate and strong $\wD$. $\Dapar$ is the frequency difference between the nuclear spins' $\apar$ coupling constants.
	}
	\label{fig3}
\end{figure*}

Figs.~\ref{fig2}(c,e) show a representative 2D spectrum for NV-A and corresponding numerical simulation.  The horizontal axis represents the direct dimension (frequency $f_1$), while the vertical axis is the indirect dimension (frequency $f_2$) encoding the Rabi driving.
In the horizontal direction, peaks are distributed between $1960-1970\unit{kHz}$.  Here, peak frequencies encode the parallel hyperfine coupling $\apar$ that shifts resonances away from the expected \C frequency (no couplings present) at $f_0 = \yn B_0/(2\pi) \approx 1965\unit{kHz}$.
The vertical direction corresponds to the Rabi driving amplitude $f_2 = \yn B_1/(4\pi)$, which is determined by the magnitude of the transverse hyperfine coupling $\aperp$.
Isolated peaks with $f_2>5\unit{kHz}$ reflect proximal \C with relative strong hyperfine interactions, while peaks with $f_2\lesssim 5\unit{kHz}$ correspond to more distant \C.

In the following, we focus our analysis on the five main peaks of Fig.~\ref{fig2}(c).  From the frequency coordinates $(f_1,f_2)$ of a peak we extract the hyperfine coupling parameters ($\apar$,$\aperp$) according to~\cite{boss16},
\begin{align}
\apar &= 4\pi f_1-2\yn B_0 \\
\aperp &= \pi^2 f_2  \text{,}
\label{eq:freq_convert}
\end{align}
using the approximation $\yn B_0 \gg \apar,\aperp$.  The ($\apar$,$\aperp$) values may be converted to spatial position vectors by including the phase information (SM and Ref.~\cite{cujia22}).  To corroborate this procedure, we perform an \textit{ab initio} density matrix simulation of a five-spin cluster, taking extracted spatial position vectors as input and accounting for dipolar interactions between nuclei (Fig.~\ref{fig2}(e)).
Excellent agreement to experimental peak positions are obtained with modest agreement to measured peak intensities. Variation in latter may be due to inconsistent polarization of  nuclei.
For comparison, Figs.~\ref{fig2}(d,f) show corresponding 1D spectra, where the overlap between nearby \C peaks with similar $\apar$ complicates spectral analysis.

As a second application, we demonstrate identification of nuclear spin pairs.
To detect spin pairs, we use a scheme derived from $J$-spectroscopy~\cite{aue76} (Fig.~\ref{fig3}(b)). The manipulation step of Fig.~\ref{fig1}(c) is replaced by a nuclear spin echo, and we increment total echo duration $t_2$.  The spin echo eliminates static Zeeman and parallel hyperfine terms of the nuclear-spin Hamiltonian, such that evolution during $t_2$ is only due to nuclear spin-spin interactions. For two spins $i$ and $j$ with through-space distance $r_{ij}$ forming an angle $\theta_{ij}$ with the external bias field, coupling frequency is given by
\begin{equation}
\wD = \pm \frac{3}{4} \frac{\mu_0 \hbar}{4 \pi} \frac{\yn^2}{r_{ij}^3} (1-3 \cos^2 \theta_{ij}) .
\label{eq:wD}
\end{equation}
Compared to conventional $J$-spectroscopy, dipolar coupling dominates over scalar ($J$) coupling, while the spin Hamiltonian takes the same form except for different coupling constants.
Acquiring a 2D dataset and applying a Fourier transform with respect to $t_1$ and $t_2$, the coupling frequency $\wD = 2\pi f_2$ of a spin pair can be extracted along the indirect dimension, and correlated with peak positions in the 1D spectrum.

Figs.~\ref{fig3}(c-f) show 2D and projected 1D spectra of the \C environment of NV-B and NV-C.  Most peaks lie near the $f_2 \sim 0\unit{Hz}$ center line, signaling isolated \C nuclei. A few satellite peaks with comparably large $|f_2|\gtrsim 200\unit{Hz}$ are visible in both spectra.  Specifically, satellite peaks appear at $\pm 1030\unit{Hz}$ for NV-B (Fig.~\ref{fig3}(d)) and at $\pm 200\unit{Hz}$ for NV-C (Fig.~\ref{fig3}(e)).  Since the diamond lattice only permits discrete values of coupling frequencies corresponding to a specific \C configurations, we use the measured $f_2$ to identify spin pairs.  
We assign the satellite peaks to three spin pairs: a nearest-neighbor pair along the $(1\bar{1}\bar{1})$ crystal axis (NV-B), and two next-next-nearest neighbor pairs along $(\bar{1}13)$ (NV-C) (Figs.~\ref{fig3}(c,d) insets).

To gain insight into the multiplet structure, we simulate the expected peak pattern for a spin pair with weak ($\wD\ll\Dapar$), intermediate ($\wD\approx\Dapar$) and strong ($\wD\gg\Dapar$) coupling, where $\Dapar$ is the difference between the nuclear spin's $\apar$ parameter.  Qualitative behavior is shown in Fig.~\ref{fig3}(g) and quantitative results are given in the SM~\cite{supplemental}.  In the weak-coupling case, each \C resonance is split into a peak doublet separated by $2\wD$ (left).  As $\wD$ becomes comparable to $\Dapar$ (middle), the peaks are more evenly distributed and frequencies are not straightforward to interpret~\cite{supplemental}.  Once $\wD\gg\ \Dapar$ (right), only two peaks are observed, lying near the diagonal and separated by $2\wD$.

Revisiting the experimental spectra in Fig.~\ref{fig3}(c,d), we qualitatively identify pair (1) near NV-B as a strong-coupling case, and pairs (2) and (2') near NV-C as weak-to-intermediate coupling cases.  However, only three out of four peaks (pair (2)) resp. two out for four peaks (pair (2')) are visible in Fig.~\ref{fig3}(d).  This is because peak intensities are in general different~\cite{supplemental} and since the signal-to-noise ratio is limited, some peaks fall below the detection threshold.

Both 2D spectra show structure near the $f_2=0\unit{Hz}$ center line, suggesting that more spin pairs are present. However, our spectral resolution ($\sim 100\unit{Hz}$) is insufficient to assign these peaks to specific pairs of lattice sites.

Unlike experiments conducted at cryogenic temperatures, where sub-Hz \C resolution can be reached~\cite{abobeih19}, our spectral resolution is limited by coherence times of the \C spins ($T_{2,\mr{n}} \sim 10\unit{ms}$), bounded by the room-temperature $T_1$ time of the NV~\cite{rosskopf17}.

Future work will therefore need to focus on improving the spectral resolution.  Unlike experiments conducted at cryogenic temperatures, where sub-Hz \C resolution can be reached~\cite{abobeih19}, extending the coherence at room temperature is not straightforward.  In our experiments, the coherence time of the \C spins ($T_{2,\mr{n}} \sim 10\unit{ms}$) (and thus, the spectral resolution) is bounded by the room-temperature $T_1$ time of the NV and by bias field drifts~\cite{rosskopf17}.
One approach to prolonging $T_{2,\mr{n}}$ is the use of dual-frequency driving similar to Fig.~\ref{fig2}(b), where the weak measurement readout alternates between the two NV transitions combined with a periodic inversion of the nuclear spin.  This approach requires higher $\mu$w driving powers than currently available with our setup, but is the subject of future work.  Alternatively, optical control of the NV spin and charge state~\cite{maurer12,pfender17nl} could be explored to extend the \C coherence.

In conclusion, we demonstrate a set of methods for extending single-spin NMR detection to multiple spectral dimensions.  Our approach combines weak-measurement spectroscopy~\cite{cujia19} with an additional, indirect evolution period well-known from multidimensional NMR spectroscopy~\cite{aue76}. We demonstrate precise measurement of the transverse component of the NV-\C hyperfine interaction, and selective editing of nuclear spin pairs including quantification of dipolar-coupling constants. The 2D spectra reduce spectral overlap and enable improved assignment of single \C peaks.

Although the number of assigned peaks in this work (\eg, $n=5$ in Fig.~\ref{fig2}(c)) is smaller compared to previous work where $n>20$ \cite{cujia22,abobeih19,stolpe23}, we consider multi-dimensional techniques as an essential ingredient towards improved localization of intrinsic and extrinsic nuclear spins.  To extend analysis to larger spin clusters, multidimensional cross-correlation methods derived from COSY or NOESY spectroscopy~\cite{keeler11} could be adapted or higher spectral dimensions with additional indirect evolution periods included.
Assuming challenges in stabilizing near-surface NV centers~\cite{sangtawesin19,janitz22} and molecular functionalization of diamond~\cite{liu22,xie22,abendroth22} can be overcome, multidimensional approaches as presented here will add an important tool for spectroscopic analysis of surface-anchored molecules.

The authors thank Jonathan Zopes for fruitful discussions.  This work has been supported by Swiss National Science Foundation (SNFS) Project Grant No. 200020-175600, the National Center of Competence in Research in Quantum Science and Technology (NCCR QSIT), and the Advancing Science and TEchnology thRough dIamond Quantum Sensing (ASTERIQS) program, Grant No. 820394, of the European Commission.  EJ acknowledges support from a Natural Sciences and Engineering Research Council of Canada (NSERC) postdoctoral fellowship (PDF-558200-2021).

\bibliography{library}

\end{document}